\documentclass[11pt,a4paper]{article}
\usepackage{amssymb}
\usepackage{amsmath}
\usepackage{amsthm}
\usepackage{array}
\begin{document}

\title{\Large A Generalization of Cyclic Code and Applications to Public Key Cryptosystems}

\author{Zheng Zhiyong, Huang Wenlin, Xu Jie and Tian Kun$^{*}$
\\ \small{Engineering Research Center of Ministry of Education for Financial Computing}
\\ \small{and Digital Engineering, Renmin University of China, Beijing, 100872, China}
\\ \small{$^{*}$Corresponding author email: tkun19891208@ruc.edu.cn}}
\date{}
\maketitle

\begin{abstract}

In this paper, we define and discuss $\phi$-cyclic code, which may be regarded as a general form of the ordinary cyclic code. As applications, we explain how to extend two public key encryption schemes, one is McEliece and Niederriter's cryptosystem of which is based on error-correcting code theory. Another one is NTRU public key cryptosystem of which is based on polynomial ring theory. The main purpose of this paper is to give a more general construction of NTRU based on ideal matrices and $q$-ary lattice theory.

\end{abstract}


\newpage
\numberwithin{equation}{section}
\section{$\phi$-Cyclic Code}

Let $F_q$ be a finite field with $q$ elements and $q$ be a power of a prime number, $F_q[x]$ be the polynomial ring of $F_q$ with variable $x$. Let $F_q^n$ be the $n$-dimensional linear space over $F_q$, and $a=(a_0,a_1,...,a_{n-1})\in F_q^n$ be a fixed vector in $F_q^n$ with $a_0\neq 0$, the associated polynomial of $a$ given by
\begin{equation}
\phi(x)=\phi_a(x)=x^n-a_{n-1}x^{n-1}-\cdots-a_1 x-a_0\in F_q[x], a_0\neq 0.
\end{equation}
Let $<\phi(x)>$ be the principal ideal generated by $\phi(x)$ in $F_q[x]$. There is a one to one correspondence between $F_q^n$ and the quotient ring $R=F_q[x] / <\phi(x)>$, given by
\begin{equation}
c=(c_0,c_1,...,c_{n-1})\in F_q^n\rightleftarrows c(x)=c_0+c_1 x+\cdots+c_{n-1}x^{n-1}\in R.
\end{equation}
In fact, this correspondence is also an isomorphism of Abel groups. One may extend this correspondence to subsets of $F_q^n$ and $R$ by
\begin{equation}
C\subset F_q^n \rightleftarrows C(x)=\{c(x)|c\in C\}\subset R.
\end{equation}
If $C\subset F_q^n$ is a linear subspace of $F_q^n$ of dimension $k$, then $C$ is called a linear code in coding theory and written by $C=[n,k]$ as usual. Each vector $c=(c_0,c_1,...,c_{n-1})\in C$ is called a codeword of length $n$. Obviously, $C=[n,0]$ and $C=[n,n]$ are two trivial codes. Another one is called constant codes, of which is almost trivial given by
\begin{equation*}
C=\{(b,b,...,b)|b\in F_q\}, \text{and}\ C=[n,1].
\end{equation*}
According to the given polynomial $\phi(x)=\phi_a(x)$, we may define a linear transformation $\tau_{\phi}$ in $F_q^n$,
\begin{equation}
\tau_{\phi}(c)=\tau_{\phi}((c_0,c_1,...,c_{n-1}))=(a_0 c_{n-1},c_0+a_1 c_{n-1},c_1+a_2 c_{n-1},...,c_{n-2}+a_{n-1}c_{n-1}).
\end{equation}
It is easily seen that $\tau_{\phi}:F_q^n\rightarrow F_q^n$ is a linear transformation.\\

\textbf{Definition 1.1} Let $C\subset F_q^n$ be a linear code. It is called a $\phi$-cyclic code, if
\begin{equation}
\forall c\in C\Rightarrow \tau_{\phi}(c)\in C.
\end{equation}
In other words, a linear code $C$ is a $\phi$-cyclic code, if and only if $C$ is closed under linear transformation $\tau_{\phi}$. Clearly, if $a=(1,0,...,0)$, and $\phi_a(x)=x^n-1$, then the $\phi$-cyclic code is precisely the ordinary cyclic code (see Chapter 6 of [9]).

We first show that there is a one to one correspondence between $\phi$-cyclic codes in $F_q^n$ and ideals in $R=F_q[x] / <\phi(x)>$.

\textbf{Theorem 1} Let $C\subset F_q^n$ be a subset, then $C$ is a $\phi$-cyclic code, if and only if $C(x)$ is an ideal of $R$.

\textbf{Proof:} We use column notation for vector in $F_q^n$, then linear transformation $\tau_{\phi}$ may be written as
\begin{equation*}
\tau_{\phi}
\begin{pmatrix}
     c_0 \\
     c_1 \\
     \vdots \\
     c_{n-1}
\end{pmatrix}
=
\begin{pmatrix}
     a_0 c_{n-1} \\
     c_0+a_1 c_{n-1} \\
     \vdots \\
     c_{n-2}+a_{n-1}c_{n-1}
\end{pmatrix},
\forall c=
\begin{pmatrix}
     c_0 \\
     c_1 \\
     \vdots \\
     c_{n-1}
\end{pmatrix}
\in F_q^n.
\end{equation*}
Let $T_{\phi}$ be a $n\times n$ square matrix over $F_q$,
\begin{equation}
T_{\phi}=
\left(
\begin{array}{ccc|c}
0 & \cdots & 0 & a_0\\
\hline
& & & a_1\\
& I_{n-1} & & \vdots \\
& & & a_{n-1} \\
\end{array}
\right)
\in F_q^{n\times n}.
\end{equation}
where $I_{n-1}$ is the $(n-1)\times(n-1)$ unit matrix. The matrix expression of $\tau_{\phi}$ as follows
\begin{equation}
\tau_{\phi}
\begin{pmatrix}
     c_0 \\
     c_1 \\
     \vdots \\
     c_{n-1}
\end{pmatrix}
=
T_{\phi}
\begin{pmatrix}
     c_0 \\
     c_1 \\
     \vdots \\
     c_{n-1}
\end{pmatrix}
=
\begin{pmatrix}
     a_0 c_{n-1} \\
     c_0+a_1 c_{n-1} \\
     \vdots \\
     c_{n-2}+a_{n-1}c_{n-1}.
\end{pmatrix}.
\end{equation}
Suppose $C\subset F_q^n$ and $C(x)$ is an ideal of $R$, it is clear that $C$ is a linear code of $F_q^n$. To prove $C$ is a $\phi$-cyclic code, we note that for any polynomial $c(x)\in C(x)$, then $xc(x)\in C(x)$ if and only if $\tau_{\phi}(c)\in C$, namely, if $c(x)\in C(x)$, then
\begin{equation}
xc(x)\in C(x) \Leftrightarrow \tau_{\phi}(c)\in C \Leftrightarrow T_{\phi} c\in C.
\end{equation}
Therefore, if $C(x)$ is an ideal of $R$, then we have immediately that $C$ is a $\phi$-cyclic code of $F_q^n$.

Conversely, if $C\subset F_q^n$ is a $\phi$-cyclic code, then for all $k\geqslant 1$, we have
\begin{equation*}
\forall c\in C \Rightarrow T_{\phi}^k c\in C, k\geqslant 1.
\end{equation*}
It follows that
\begin{equation*}
\forall c(x)\in C(x) \Rightarrow x^k c(x)\in C(x), 0\leqslant k \leqslant n-1,
\end{equation*}
which implies $C(x)$ is an ideal of $R$. This is the proof of Theorem 1.\qquad $\Box$

By Theorem 1, to find a $\phi$-cyclic code, it is enough to find an ideal of $R$. There are two trivial ideals $C(x)=0$ and $C(x)=R$, the corresponding $\phi$-cyclic codes are $C=[n,0]$ and $C=F_q^n$ respectively, which are called trivial $\phi$-cyclic code. To find non-trivial $\phi$-cyclic codes, we make use of homomorphic theorems, which is a standard technique in Algebra. Let $\pi$ be the natural homomorphism from $F_q[x]$ to its quotient ring $R=F_q[x] / <\phi(x)>$, $\text{ker}\pi=<\phi(x)>$,
\begin{equation}
<\phi(x)>\subset N \subset F_q[x] \xrightarrow{\quad\pi\quad} R=F_q[x] / <\phi(x)>,
\end{equation}
where $N$ is an ideal of $F_q[x]$, of which is containing $\text{ker}\pi=<\phi(x)>$. Since $F_q[x]$ is a principal ideal domain, then $N=<g(x)>$ is a principal ideal generated by a monic polynomial $g(x)\in F_q[x]$. It is easy to see that
\begin{equation*}
<\phi(x)>\subset <g(x)> \Leftrightarrow g(x)|\phi(x).
\end{equation*}
It follows that all ideals $N$ satisfing (1.9) are given by
\begin{equation*}
\{<g(x)>\ |\ g(x)\in F_q[x] \text{\ is monic and\ } g(x)|\phi(x)\}.
\end{equation*}
We write by $<g(x)>$ mod $\phi(x)$, the image of $<g(x)>$ under $\pi$, it is easy to check
\begin{equation}
<g(x)> \text{\ mod\ } \phi(x)=\{h(x)g(x)\ |\ h(x)\in F_q[x] \text{\ and deg} h(x)+\text{deg}g(x)<n\},
\end{equation}
more precisely, which is a representative elements set of $<g(x)>$ mod $\phi(x)$, by homomorphism theorem in ring theory, all ideals of $R$ given by
\begin{equation}
\{<g(x)> \text{\ mod\ } \phi(x) \ |\ g(x) \in F_q[x] \text{\ is monic and\ } g(x)|\phi(x)\}.
\end{equation}
Let $d$ be the number of monic divisors of $\phi(x)$ in $F_q[x]$, it follows immediately that

\textbf{Corollary 1} The number of $\phi$-cyclic code in $F_q^n$ is $d$.

To compare the $\phi$-cyclic code and ordinary cyclic code, we see a simple example.

\textbf{Example 1} Constant code $C$ is always a cyclic code for $1+x+\cdots+x^{n-1}|x^n-1$, and its generated polynomial is just $1+x+\cdots+x^{n-1}$. But constant code $C$ in $F_q^n$ is not always a $\phi$-cyclic code, it is a $\phi$-cyclic code if and only if $1+x+\cdots+x^{n-1}|\phi(x)$, an equivalent condition for $1+x+\cdots+x^{n-1}|\phi(x)$ is
\begin{equation*}
a_{n-1}=a_{n-2}=\cdots=a_1=b,\text{\ and\ } a_0=1+b.
\end{equation*}

\textbf{Definition 1.2} Let $C$ be a $\phi$-cyclic code and $C(x)=g(x)$ mod $\phi(x)$. We call $g(x)$ is the generated polynomial of $C$, where $g(x)$ is monic and $g(x)|\phi(x)$.

\textbf{Lemma 1.1} Let $g(x)=g_0+g_1 x+\cdots+g_{n-k-1}x^{n-k-1}+x^{n-k}$ be the generated polynomial of a $\phi$-cyclic code $C$, where $1\leqslant k\leqslant n-1$, and $g(x)|\phi(x)$, then $C=[n,k]$ and a generated matrix for $C$ is the following block matrix
\begin{equation}
G=
\begin{pmatrix}
     g \\
     \tau_{\phi}(g) \\
     \tau_{\phi}^2(g) \\
     \vdots \\
     \tau_{\phi}^{k-1}(g)
\end{pmatrix}_{k\times n},
\end{equation}
where $g=(g_0,g_1,...,g_{n-k-1},1,0,...,0)\in C$ is the corresponding codeword of $g(x)$, and $\tau_{\phi}^{i}(g)=\tau_{\phi}^{i-1}(\tau_{\phi}(g))$ for $1\leqslant i \leqslant n-1$.

\textbf{Proof:} By assumption, $C(x)=<g(x)>$ mod $\phi(x)$, then $\{g,\tau_{\phi}(g),...,\tau_{\phi}^{k-1}(g)\}\subset C$, we are to prove it is a basis of $C$. First, these vectors are linearly independent. Otherwise, we have
\begin{equation}
\sum_{i=0}^{k-1} b_i \tau_{\phi}^i (g)=0, \text{\ for some\ } b_i\in F_q,
\end{equation}
and the corresponding polynomial is zero, namely
\begin{equation*}
(\sum_{i=0}^{k-1} b_i x^i)g(x)=0.
\end{equation*}
It follows that
\begin{equation*}
\sum_{i=0}^{k-1} b_i x^i=0\Rightarrow b_i=0 \text{\ for all\ } i, 0\leqslant i\leqslant k-1.
\end{equation*}
Next, if $c\in C$, and $c(x)\in C(x)$, by (1.10), there is a polynomial $b(x)=b_0+b_1 x+\cdots+b_{k-2}x^{k-2}+x^{k-1}$ such that
\begin{equation*}
c(x)=b(x)g(x)=(\sum_{i=0}^{k-1} b_i x^i)g(x), \text{\ where\ } b_{k-1}=1.
\end{equation*}
Thus we have the corresponding codeword of $C(x)$
\begin{equation*}
c=\sum_{i=0}^{k-1} b_i \tau_{\phi}^i (g).
\end{equation*}
This shows that $\{g,\tau_{\phi}(g),...,\tau_{\phi}^{k-1}(g)\}$ is a basis of $C$, and a generated matrix for $C$ is
\begin{equation*}
G=
\begin{pmatrix}
     g \\
     \tau_{\phi}(g) \\
     \tau_{\phi}^2(g) \\
     \vdots \\
     \tau_{\phi}^{k-1}(g)
\end{pmatrix}_{k\times n}.
\end{equation*}
We have lemma 1.1 at once.\qquad\qquad\qquad\qquad\qquad\qquad\qquad\qquad\qquad\qquad $\Box$

To describe a parity check matrix for a $\phi$-cyclic code, for any $c=(c_0,c_1,...,c_{n-1})\in F_q^n$, we write
\begin{equation*}
\overline{c}=(c_{n-1},c_{n-2},...,c_1,c_0)\in F_q^n.
\end{equation*}

\textbf{Lemma 1.2} Suppose $C$ is a $\phi$-cyclic code with generated polynomial $g(x)$, where $g(x)|\phi(x)$ and $\text{deg}g(x)=n-k$. Let $h(x)g(x)=\phi(x)$, where $h(x)=h_0+h_1 x+\cdots+h_{k-1}x^{k-1}+x^k$. Then a parity check matrix for $C$ is
\begin{equation}
H=
\begin{pmatrix}
     \overline{h} \\
     \tau_{\phi}(\overline{h}) \\
     \vdots \\
     \tau_{\phi}^{n-k-1}(\overline{h})
\end{pmatrix}_{(n-k)\times n}.
\end{equation}

\textbf{Proof:} Since $h(x)g(x)=\phi(x)$, it means that $h(x)g(x)=0$ in $R=F_q[x]/<\phi(x)>$, thus we have
\begin{equation*}
g_0 h_i+g_1 h_{i-1}+\cdots+g_{n-k}h_{i-n+k}=0,\forall 0\leqslant i\leqslant n-1.
\end{equation*}
It follows that $GH'=0$, where $G$ is a generated matrix for $C$ given by (1.12). Therefore, $H$ is a parity check matrix for $C$.\qquad\qquad\qquad\qquad\qquad\qquad $\Box$

A separable polynomial in Algebra means that it has no multiple roots in its splitting field. The following lemma shows that there is an unit element in any non-zero ideal of $R$, when $\phi(x)$ is a separable polynomial.

\textbf{Lemma 1.3} Suppose $\phi(x)$ is a separable polynomial of $F_q$, and $C(x)=g(x)$ mod $\phi(x)$ is an ideal of $R$ with $\text{deg}g(x)\leqslant n-1$, then there exists an element $d(x)\in C(x)$ such that
\begin{equation*}
c(x)d(x)=c(x),\text{\ for all\ } c(x)\in C(x).
\end{equation*}

\textbf{Proof:} Let $h(x)g(x)=\phi(x)$. Since $\phi(x)$ is a separable polynomial, then $\text{gcd}(g(x),h(x))=1$, and there are two polynomial $a(x)$ and $b(x)$ in $F_q[x]$ such that
\begin{equation*}
a(x)g(x)+b(x)h(x)=1.
\end{equation*}
Let
\begin{equation*}
d(x)=a(x)g(x)=1-b(x)h(x)\in C(x).
\end{equation*}
If $c(x)\in C(x)$, by (1.10), we write $c(x)=g(x)g_1(x)$, it follows that
\begin{equation*}
c(x)d(x)\equiv a(x)g(x)g(x)g_1(x)\equiv (1-b(x)h(x))g(x)g_1(x)
\end{equation*}
\begin{equation*}
\equiv g(x)g_1(x)\equiv c(x) (\text{mod\ } \phi(x)).
\end{equation*}
Thus we have $c(x)d(x)=c(x)$ in $R$.\qquad\qquad\qquad\qquad\qquad\qquad\qquad\qquad $\Box$

Next, we discuss maximal $\phi$-cyclic code. Let $C(x)=g(x)$ mod $\phi(x)$, and $g(x)$ be an irreducible polynomial in $F_q[x]$, we call the corresponding $\phi$-cyclic code $C$ a maximal $\phi$-cyclic code, because $<g(x)>$ is a maximal ideal in $F_q[x]$.

\textbf{Lemma 1.4} Let $C$ be a maximal $\phi$-cyclic code with generated polynomial $g(x)$, $\beta$ be a root of $g(x)$ in some extensions of $F_q$, then
\begin{equation}
C(x)=\{a(x)\ |\ a(x)\in R \text{\ and\ } a(\beta)=0\}.
\end{equation}

\textbf{Proof:} If $a(x)\in C(x)$, by (1.10) we have $a(\beta)=0$ immediately. Conversely, if $a(x)\in F_q[x]$ and $a(\beta)=0$, since $g(x)$ is irreducible, thus we have $g(x)|a(x)$, and (1.15) follows at once.\qquad\qquad\qquad\qquad\qquad\qquad\qquad\qquad $\Box$

An important application of maximal $\phi$-cyclic code is to constract an error-correcting code, so that we may obtain a modified McEliece-Niederriter's cryptosystem. To do this, let $1\leqslant m<\sqrt{n}$, and $F_{q^m}$ be an extension field of $F_q$ of degree $m$. Suppose $F_{q^m}=F_q(\theta)$, where $\theta$ is a primitive element of $F_{q^m}$ and $F_q(\theta)$ is the simple extension containing $F_q$ and $\theta$. Let $g(x)\in F_q[x]$ be the minimum polynomial of $\theta$, then $g(x)$ is an irreducible polynomial of degree $m$ of $F_q[x]$. It is well-known that $F_{q^m}$ is a Galois extension of $F_q$, so that all roots of $g(x)$ are in $F_{q^m}$. Let $\beta_1,\beta_2,...,\beta_m$ be all roots of $g(x)$, the Vandermonde matrix $V(\beta_1,\beta_2,...,\beta_n)$ defined by
\begin{equation}
H=V(\beta_1,\beta_2,...,\beta_n)=
\begin{pmatrix}
1 & \beta_1 & \beta_1^2 & \cdots & \beta_1^{n-1} \\
1 & \beta_2 & \beta_2^2 & \cdots & \beta_2^{n-1}  \\
\vdots & \vdots & \vdots &  & \vdots \\
1 & \beta_m & \beta_m^2 & \cdots & \beta_m^{n-1}
\end{pmatrix}_{m\times n},
\end{equation}
where $\beta_1=\theta$ and each $\beta_i$ is a vector of $(F_q)^m$. For arbitrary monic polynomial $h(x)\in F_q[x]$, $\text{deg}h(x)=n-m$, let $\phi(x)=h(x)g(x)$ and $C$ be a maximal $\phi$-cyclic code generated by $g(x)$. It is easy to verify that
\begin{equation*}
c\in C\Leftrightarrow cH'=0.
\end{equation*}
Therefore, $H$ is a parity check matrix for $C$. If we choose the primitive element $\theta$, so that any $d-1$ columns in $H$ are linearly independent, then the minimum distance of $C$ is greater than $d$, and $C$ is a t-error-correcting code, where $t=[\frac{d}{2}]$.

The public key cryptosystems based on algebraic coding theory were created by R.J.McEliece [11] and H.Niederriter [14], a suitable t-error-correcting code plays a key role in their construction. The error-correcting code $C$ should satisfy the following requirements:

(i) $C$ should have a relatively large error-correcting capability so that a reasonable number of message vectors can be used;

(ii) $C$ should allow an efficient decoding algorithm so that the decryption can be carried out with a short time.

Our results supply a different way to choose an error-correcting code by selecting arbitrary irreducible polynomials $g(x)\in F_q[x]$ of degree $m$ and roots of $g(x)$ rather than an irreducible factor of $x^n-1$ and the roots of unit such as ordinary BCH code and Gappa code.

In fact, for any positive integer $m$, there is at least an irreducible polynomial $g(x)\in F_q[x]$ with degree $m$. Let $N_q(m)$ be the number of irreducible polynomials of degree $m$ in $F_q[x]$, then we have (see Theorem 3.25 of [8])
\begin{equation*}
N_q(m)=\frac{1}{m} \sum_{d|m} u(\frac{m}{d})q^d=\frac{1}{m} \sum_{d|m} u(d)q^{\frac{m}{d}},
\end{equation*}
where $u(d)$ is Mobi$\ddot{u}$s function.

Assuming one has selected two monic and irreducible polynomials $g(x)$ and $h(x)$ with $\text{deg}g(x)=m$ and $\text{deg}h(x)=n-m$, let $\phi(x)=g(x)h(x)$, then one may obtain $\phi$-cyclic code $C$ generated by $g(x)$ or $h(x)$, which is more convenient and more flexible than the ordinary methods.

\section{A Generalization of NTRUEncrypt}

The public key cryptosystem NTRU proposed in 1996 by Hoffstein, Pipher and Silverman, is the fastest known lattice based encryption scheme, although its description relies on arithmetic over polynomial quotient ring $Z[x]/<x^n-1>$, it was easily observed that it could be expressed as a lattice based cryptosystem (see [7]). For the background materials, we refer to [3], [5], [6], [10], [12] and [13]. Our strategy in this section is to replace $Z[x]/<x^n-1>$ by more general polynomial ring $Z[x]/<\phi(x)>$ and obtain a generalization of NTRUEncrypt, where $\phi(x)$ is a monic polynomial of degree $n$ with integer coefficients.

In this section, we denote $\phi(x)$ and $R$ by
\begin{equation}
\phi(x)=x^n-a_{n-1}x^{n-1}-\cdots-a_1 x-a_0\in Z[x],R=Z[x]/<\phi(x)>,a_0\neq 0.
\end{equation}
Let $H_{\phi}\in Z^{n\times n}$ be a square matrix given by
\begin{equation}
H=H_{\phi}=
\left(
\begin{array}{ccc|c}
0 & \cdots & 0 & a_0\\
\hline
& & & a_1\\
& I_{n-1} & & \vdots \\
& & & a_{n-1} \\
\end{array}
\right)_{n\times n},
\end{equation}
where $I_{n-1}$ is $(n-1)\times (n-1)$ unit matrix. Obviously, $\phi(x)$ is the characteristic polynomial of $H$, and $H$ defines a linear transformation of $\mathbb{R}^n\rightarrow \mathbb{R}^n$ by $x\rightarrow Hx$, where $\mathbb{R}$ is real number field, $x$ is a column vector of $\mathbb{R}^n$. We may extend this transformation to $\mathbb{R}^{2n}$ and denote $\sigma$ by
\begin{equation}
\sigma
\begin{pmatrix}
     \alpha \\
     \beta
\end{pmatrix}
=
\begin{pmatrix}
     H\alpha \\
     H\beta
\end{pmatrix},
\text{\ where\ }
\begin{pmatrix}
     \alpha \\
     \beta
\end{pmatrix}
\in \mathbb{R}^{2n}.
\end{equation}
Of course, $\sigma$ is again a linear transformation of $\mathbb{R}^{2n}\rightarrow \mathbb{R}^{2n}$.

According to [13], a $q$-ary lattice is a lattice $L$ such that $qZ^n\subset L \subset Z^n$, where $q$ is a positive integer.

\textbf{Definition 2.1} A $q$-ary lattice $L$ is called convolutional modular lattice, if $L$ is in even dimension $2n$ satisfying
\begin{equation}
\forall
\begin{pmatrix}
     \alpha \\
     \beta
\end{pmatrix}
\in L\Rightarrow
\sigma
\begin{pmatrix}
     \alpha \\
     \beta
\end{pmatrix}
=
\begin{pmatrix}
     H\alpha \\
     H\beta
\end{pmatrix}
\in L.
\end{equation}
In other words, a convolutional modular lattice is a $q$-ary lattice in even dimension and is closed under the linear transformation $\sigma$.

Recalling the secret key $\begin{pmatrix}f \\ g \end{pmatrix}$ of NTRU is a pair of polynomials of degree $n-1$, we may regard $f$ and $g$ as column vectors in $Z^n$. To obtain a convolutional modular lattice containing $\begin{pmatrix}f \\ g \end{pmatrix}$, we need some help of ideal matrices. An ideal matrix generated by a vector $f$ is defined by
\begin{equation}
H^*(f)=H_{\phi}^*(f)=[f,Hf,H^2 f,...,H^{n-1}f]_{n\times n},
\end{equation}
which is a block matrix in terms of each column $H^k f\ (0\leqslant k\leqslant n-1)$. It is easily seen that $H^*(f)$ is a generalization of the classical circulant matrices (see [4]), in fact, let $\phi(x)=x^n-1$, and $f(x)=f_0+f_1 x+\cdots+f_{n-1}x^{n-1}\in Z[x]$, the ideal matrix $H_{\phi}^*(f)$ generated by $f$ is given by
\begin{equation*}
H^*(f)=H_{\phi}^*(f)=
\begin{pmatrix}
f_0 & f_{n-1} & \cdots & f_1 \\
f_1 & f_0 & \cdots & f_2 \\
\vdots & \vdots &  & \vdots \\
f_{n-1} & f_{n-2} & \cdots & f_0
\end{pmatrix},
\phi(x)=x^n-1,
\end{equation*}
which is known as a circulant matrix. On the other hand, ideal matrix and ideal lattice play an important role in Ajtai's construction of a collision resistant Hash function, the related materials we refer to [1], [2], [10], [15], [16] and [17].

First, we have to establish some basic properties for an ideal matrix $H^*(f)$, most of them are known when $H^*(f)$ is a circulant matrix.

\textbf{Lemma 2.1} Suppose $H$ and $H^*(f)$ are given by (2.2) and (2.5) respectively, then for any $f\in \mathbb{R}^n$ we have
\begin{equation*}
H\cdot H^*(f)=H^*(f)\cdot H, \quad\forall f\in \mathbb{R}^n.
\end{equation*}

\textbf{Proof:} Since $\phi(x)=x^n-a_{n-1}x^{n-1}-\cdots-a_1 x-a_0$ is the characteristic polynomial of $H$, by Hamilton-Cayley theorem, we have
\begin{equation}
H^n=a_0 I_n+a_1 H+\cdots+a_{n-1}H^{n-1}.
\end{equation}
Let
\begin{equation*}
b=
\begin{pmatrix}
     a_1 \\
     a_2 \\
     \vdots \\
     a_{n-1}
\end{pmatrix},
\text{\ and\ } H=
\begin{pmatrix}
     0 & a_0 \\
     I_{n-1} & b
\end{pmatrix}.
\end{equation*}
By (2.5) we have
\begin{equation*}
H^*(f)H=[f,Hf,...,H^{n-1}f]
\begin{pmatrix}
     0 & a_0 \\
     I_{n-1} & b
\end{pmatrix}
\end{equation*}
\begin{equation*}
=[Hf,H^2f,...,H^{n-1}f,a_0 f+a_1 Hf+\cdots+a_{n-1}H^{n-1}f]
\end{equation*}
\begin{equation*}
=[Hf,H^2f,...,H^{n-1}f,H^nf]
\end{equation*}
\begin{equation*}
=H[f,Hf,...,H^{n-1}f]=H\cdot H^*(f).
\end{equation*}
the lemma follows.\qquad\qquad\qquad\qquad\qquad\qquad\qquad\qquad\qquad\qquad\qquad\qquad $\Box$

\textbf{Lemma 2.2} For any $f=\begin{pmatrix}
     f_0 \\
     f_1 \\
     \vdots \\
     f_{n-1}
\end{pmatrix}\in \mathbb{R}^n$ we have
\begin{equation}
H^*(f)=f_0 I_n+f_1 H+\cdots+f_{n-1}H^{n-1}.
\end{equation}

\textbf{Proof:} We use induction on $n$ to show this conclusion. If $n=1$, it is trivial. Suppose it is true for $n$, we consider the case of $n+1$. For this purpose, we write $H=H_n$, $e_1,e_2,...,e_n$ the $n$ column vectors of unit in $\mathbb{R}^n$, namely
\begin{equation*}
e_1=
\begin{pmatrix}
     1 \\
     0 \\
     \vdots \\
     0
\end{pmatrix},
e_2=
\begin{pmatrix}
     0 \\
     1 \\
     \vdots \\
     0
\end{pmatrix}\cdots
e_n=
\begin{pmatrix}
     0 \\
     0 \\
     \vdots \\
     1
\end{pmatrix},
\end{equation*}
and
\begin{equation*}
H_{n+1}=
\begin{pmatrix}
     0 & A_0 \\
     e_1 & Hn
\end{pmatrix},
\end{equation*}
where $A_0=(0,0,...,a_0)\in \mathbb{R}^n$ is a row vector. For any $k$, $1\leqslant k\leqslant n-1$, it is easy to check that
\begin{equation*}
H_n e_k=e_{k+1}, H_n^k e_1=e_{k+1} \text{\ and\ } H_{n+1}^k=
\begin{pmatrix}
     0 & A_0 H_n^{k-1} \\
     e_k & H_n^k
\end{pmatrix}.
\end{equation*}
Let $f=\begin{pmatrix}
     f_0 \\
     f_1 \\
     \vdots \\
     f_{n-1} \\
     f_n
\end{pmatrix}\in \mathbb{R}^{n+1}$, we denote $f'$ by
\begin{equation*}
f'=\begin{pmatrix}
     f_1 \\
     f_2 \\
     \vdots \\
     f_n
\end{pmatrix}
\in \mathbb{R}^n, \text{\ and\ }
f=\begin{pmatrix}
f_0 \\
f'
\end{pmatrix}.
\end{equation*}
By the assumption of induction, we have
\begin{equation*}
H_n^*(f')=[f',H_nf',...,H_n^{n-1}f']=f_1 I_n+f_2 H_n+\cdots+f_n H_n^{n-1}.
\end{equation*}
It follows that
\begin{equation*}
H_{n+1}^*(f)=\Big[
\begin{pmatrix}
f_0 \\
f'
\end{pmatrix},
H_{n+1}\begin{pmatrix}
f_0 \\
f'
\end{pmatrix},\cdots,H_{n+1}^n
\begin{pmatrix}
f_0 \\
f'
\end{pmatrix}
\Big]
\end{equation*}
\begin{equation*}
=f_0 I_n+f_1 H_{n+1}+\cdots+f_n H_{n+1}^n.
\end{equation*}
We complete the proof of lamma 2.2.\qquad\qquad\qquad\qquad\qquad\qquad\qquad\qquad $\Box$

We always suppose that $\phi(x)\in Z[x]$ is a separable polynomial and $w_1,w_2,...,w_n$ are complex number roots of $\phi(x)$, of which are different from each other. The Vandermonde matrix $V_{\phi}$ generated by $\{w_1,w_2,...,w_n\}$ is
\begin{equation*}
V_{\phi}=
\begin{pmatrix}
1 & 1 & \cdots & 1 \\
w_1 & w_2 & \cdots & w_n \\
\vdots & \vdots &  & \vdots \\
w_1^{n-1} & w_2^{n-1} & \cdots & w_n^{n-1}
\end{pmatrix},\quad
\text{\ and\ \ \ det}(V_{\phi})\neq 0.
\end{equation*}

\textbf{Lemma 2.3} Let $f(x)=f_0+f_1 x+\cdots+f_{n-1}x^{n-1}\in \mathbb{R}[x]$, then we have
\begin{equation}
H^*(f)=V_{\phi}^{-1}\text{\ diag\ }\{f(w_1),f(w_2),...,f(w_n)\}V_{\phi},
\end{equation}
where diag $\{f(w_1),f(w_2),...,f(w_n)\}$ is the diagonal matrix.

\textbf{Proof:} By Theorem 3.2.5 of [4], for $H$, we have
\begin{equation}
H=V_{\phi}^{-1}\text{\ diag\ }\{w_1,w_2,...,w_n\}V_{\phi}.
\end{equation}
By lemma 2.2, it follows that
\begin{equation*}
H^*(f)=V_{\phi}^{-1}\text{\ diag\ }\{f(w_1),f(w_2),...,f(w_n)\}V_{\phi}\quad.
\end{equation*}  \qquad\qquad\qquad\qquad\qquad\qquad\qquad\qquad\qquad\qquad\qquad\qquad\qquad\qquad\qquad\qquad $\Box$

Now, we summarize some basic properties for ideal matrix as follows.

\textbf{Theorem 2} Let $f\in \mathbb{R}^n$, $g\in \mathbb{R}^n$ be two column vectors and $H^*(f)$ be the ideal matrix generated by $f$, then we have:

(i) $H^*(f)H^*(g)=H^*(g)H^*(f)$.

(ii) $H^*(f)H^*(g)=H^*(H^*(f)g)$.

(iii) det $(H^*(f))=\Pi_{i=1}^n f(w_i)$.

(iv) $H^*(f)$ is an invertible matrix if and only if $\phi(x)$ and $f(x)$ are coprime, i.e. gcd $(\phi(x),f(x))=1$.

\textbf{Proof:} (i) and (ii) follow from lemma 2.2 immediately, (iii) and (iv) follow from lemma 2.3. Here we only give an equivalent form of (ii). Let
\begin{equation}
f\ast g=H^*(f) g.
\end{equation}
then by (ii) we have
\begin{equation}
H^*(f\ast g)=H^*(f)H^*(g).
\end{equation} \qquad\qquad\qquad\qquad\qquad\qquad\qquad\qquad\qquad\qquad\qquad\qquad\qquad\qquad\qquad\qquad $\Box$

To construct a convolutional modular lattice containing vector $\begin{pmatrix} f\\g \end{pmatrix}$, let $\begin{pmatrix} f\\g \end{pmatrix}\in Z^{2n}$, $(H^*(f))'$ be the transpose of $H^*(f)$, and
\begin{equation}
A=[(H^*(f))',(H^*(g))']=
\begin{pmatrix}
f' & g' \\
f'H' & g'H' \\
f'(H')^2 & g'(H')^2 \\
\vdots & \vdots \\
f'(H')^{n-1} & g'(H')^{n-1}
\end{pmatrix}_{n\times {2n}},
\end{equation}
\begin{equation}
A'=
\begin{pmatrix}
H^*(f) \\
H^*(g)
\end{pmatrix}=
\begin{pmatrix}
f & Hf & \cdots & H^{n-1}f \\
g & Hg & \cdots & H^{n-1}g
\end{pmatrix}_{2n\times n}.
\end{equation}
We consider $A$ and $A'$ as matrices over $Z_q$, i.e. $A\in Z_q^{n\times 2n}$, $A'\in Z_q^{2n\times n}$, a $q$-ary lattice $\wedge_q(A)$ is defined by (see [13])
\begin{equation}
{\wedge}_{q}(A)=\{y\in Z^{2n}\ |\ \text{there exists\ }x\in Z^n \Rightarrow y\equiv A'x (\text{mod\ } q) \}.
\end{equation}
Under the above notations, we have

\textbf{Theorem 3} For any column vectors $f\in Z^n$ and $g\in Z^n$, then $\wedge_q(A)$ is a convolutional modular lattice, and $\begin{pmatrix} f\\g \end{pmatrix}\in \wedge_q(A)$.

\textbf{Proof:} It is known that $\wedge_q(A)$ is a $q$-ary lattice, i.e.
\begin{equation*}
qZ^{2n}\subset {\wedge}_q(A) \subset Z^{2n}.
\end{equation*}
We only prove that $\wedge_q(A)$ is fixed under the linear transformation $\sigma$ given by (2.4). If $y\in \wedge_q(A)$, then $y\equiv A'x (\text{mod\ }q)$ for some $x\in Z^n$, by lemma 2.1, we have
\begin{equation*}
\sigma(y)\equiv
\begin{pmatrix}
HH^*(f)x \\
HH^*(g)x
\end{pmatrix}
=
\begin{pmatrix}
H^*(f)Hx \\
H^*(g)Hx
\end{pmatrix}
\end{equation*}
\begin{equation*}
\qquad\qquad\qquad\qquad\qquad
\equiv A'Hx (\text{mod\ }q).
\end{equation*}
It means that $\sigma(y)\in \wedge_q(A)$ whenever $y\in \wedge_q(A)$. Let
\begin{equation*}
e=
\begin{pmatrix}
1 \\
0 \\
\vdots \\
0
\end{pmatrix}
\in Z^n \Rightarrow H^*(f)e=f, \text{\ and\ } H^*(g)e=g.
\end{equation*}
We have $\begin{pmatrix} f\\g \end{pmatrix}\in \wedge_q(A)$, and Theorem 3 follows.\qquad\qquad\qquad\qquad\qquad\qquad $\Box$

Since $\wedge_q(A)\subset Z^{2n}$, then there is a unique Hermite Normal Form of basis $N$, which is a upper triangular matrix given by
\begin{equation}
N=
\begin{pmatrix}
I_n & H^*(h) \\
0 & qI_n
\end{pmatrix},
\text{\ where\ } h\equiv (H^*(f))^{-1}g (\text{mod\ }q).
\end{equation}

Next, we consider parameters system of NTRU. To choose the parameters of NTRU, let $d_f$ be a positive integer and $\{p,0,-p\}^n \subset Z^n$ be a subset of $Z^n$, of which has exactly $d_f+1$ positive entries and $d_f$ negative ones, the remaining $n-2d_f-1$ entries will be zero. We take some assumption conditions for choice of parameters as follows:

(i) $\phi(x)=x^n-a_{n-1}x^{n-1}-\cdots-a_1 x-a_0\in Z[x]$ with $a_0\neq 0$, and $\phi(x)$ is separable polynomial, $n,p,q,d_f$ are positive integers with $n$ prime, $1<p<q$ and gcd $(p,q)=1$.

(ii) $f(x)$ and $g(x)$ are two polynomials in $Z[x]$ of degree $n-1$, the constant term of $f(x)$ is $1$, and
\begin{equation*}
f(x)-1\in\{p,0,-p\}^n,\quad g\in\{p,0,-p\}^n.
\end{equation*}

(iii) $H^*(f)$ is invertible modulo $q$.

(iv) $d_f<(\frac{q}{2}-1)/4p-\frac{1}{2}$.

Under the above conditions, by lemma 2.2 we have
\begin{equation}
H^*(f)\equiv I_n (\text{mod\ }p),\text{\ and\ } H^*(g)\equiv 0 (\text{mod\ }p).
\end{equation}

Now, we state a generalization of NTRU as follows.

$\bullet$ Private key. The private key in generalized NTRU is a short vector $\begin{pmatrix} f\\g \end{pmatrix}\in Z^{2n}$. The lattice associated with a private key is $\wedge_q(A)$, which is a convolutional modular lattice containing private key.

$\bullet$ Public key. The public key of the generalized NTRU is the HNF basis $N$ of $\wedge_q(A)$, which is given by (2.15).

$\bullet$ Encryption. An input message is encoded as a vector $m\in\{1,0,-1\}^n$ with exactly $d_f+1$ positive entries and $d_f$ negative ones. The vector $m$ is concatenated with a randomly chosen vector $r\in\{1,0,-1\}^n$ also with exactly $d_f+1$ positive entries and $d_f$ negative ones, to obtain a short error vector $\begin{pmatrix} m\\r \end{pmatrix}\in \{1,0,-1\}^{2n}$. Let
\begin{equation}
\begin{pmatrix}
c\\
0
\end{pmatrix}
=N
\begin{pmatrix}
m\\
r
\end{pmatrix}
\equiv
\begin{pmatrix}
m+H^*(h)r\\
0
\end{pmatrix}
(\text{mod\ }q),
\end{equation}
where $h$ is given by (2.15). Then, the $n$-dimensional vector $c$
\begin{equation*}
c\equiv m+H^*(h)r (\text{mod\ }q),
\end{equation*}
is the ciphertext.

$\bullet$ Decryption. Suppose the entries of $n$-dimensional vector $c$ are belong to interval $[-\frac{q}{2},\frac{q}{2}]$, then ciphertext $c$ is decrypted by multiplying it by the secret matrix $H^*(f)$ mod $q$, it follows that
\begin{equation}
H^*(f)c\equiv H^*(f)m+H^*(f)H^*(h)r\equiv H^*(f)m+H^*(g)r (\text{mod\ }q).
\end{equation}
Here, we use the identity (ii) of Theorem 2, namely,
\begin{equation*}
H^*(f)H^*(g)=H^*(H^*(f)g).
\end{equation*}
If the above conditions (iv) is satisfied, it is easily seen that the coordinates of vector $H^*(f)m+H^*(g)r$ are all bounded by $\frac{q}{2}$ in absolute value, or, with high probability, even for larger value of $d_f$. The decryption process is completed by reducing (2.18) modulo $p$, to obtain
\begin{equation*}
H^*(f)m+H^*(g)r\equiv m I_n (\text{mod\ }p).
\end{equation*}
Thus one gets plaintext $m$ from ciphertext $c$.

\end{document}